\documentclass[11pt]{article}
\usepackage{amssymb}
\usepackage{amsmath}
\usepackage{amscd}
\usepackage{latexsym}
\usepackage{graphicx}

\catcode`@=11
\makeatletter\renewcommand{\section}{\@startsection
{section}{1}{\z@}{-3.5ex plus -1ex minus
    -.2ex}{2.3ex plus .2ex}{\large\bf }}

\makeatletter\renewcommand{\subsection}{\@startsection{subsection}{2}{\z@}{-3.25ex
plus -1ex minus
   -.2ex}{1.5ex plus .2ex}{\bf }}

\numberwithin{equation}{section}
\newcounter{saveeqn}

\catcode`@=12
\def\bea{\begin{eqnarray}}
\def\eea{\end{eqnarray}}
\def\et{\tilde e}
\def\It{\tilde I}
\def\Th{\widetilde \Theta}
\def\Ct{\widetilde C}
\def\a{\alpha}
\def\b{\beta}
\def\g{\gamma}

\def\De{\Delta}

\def\s{\varsigma}

\def\vp{\varphi}
\def\om{\omega}
\def\1{{\bar 1}}
\def\2{{\bar 2}}
\def\3{{\bar 3}}
\def\4{{\bar 4}}
\def\ab{\bar\alpha}
\def\bb{\bar\beta}
\newcommand{\ah}{\hat{a}}
\newcommand{\bh}{\hat{b}}
\newcommand{\ch}{\hat{c}}
\newcommand{\C}{\mathbb C}
\newcommand{\R}{\mathbb R}
\newcommand{\F}{\mathbb F}
\newcommand{\Z}{\mathbb Z}

\newcommand{\Acal}{{\cal A}}

\newcommand{\Ccal}{{\cal C}}
\newcommand{\Fcal}{{\cal F}}

\newcommand{\lrc}{\mathop{\lrcorner}}
\newcommand{\hra}{\mathop{\hookrightarrow}}
\def\im{\textrm{i}}
\def\ep{\textrm{e}}

\def\diff{\textrm{d}}
\def\tr{\textrm{tr}}
\def\sfrac#1#2{{\textstyle\frac{#1}{#2}}}
\def\+{\dagger}
\def\={\ =\ }
\def\und{\qquad\textrm{and}\qquad}
\def\and{\quad\textrm{and}\quad}

\def\for{\quad\textrm{for}\quad}

\begin{document}

\begin{titlepage}
\setcounter{page}{0}
\begin{flushright}
ITP--UH--01/11\\
\end{flushright}

\vskip 2.0cm

\begin{center}

{\Large\bf
Chern-Simons flows on Aloff-Wallach spaces\\[8pt]
 and Spin(7)-instantons
}

\vspace{12mm}

{\large Alexander~S.~Haupt${}^\+$,\  Tatiana~A.~Ivanova${}^{*}$,\
Olaf~Lechtenfeld${}^{\+\times}$ \ and \ Alexander~D.~Popov${}^{*}$}
\\[8mm]
\noindent ${}^\dagger${\em
Institut f\"ur Theoretische Physik,
Leibniz Universit\"at Hannover \\
Appelstra\ss{}e 2, 30167 Hannover, Germany }\\
{Emails: Alexander.Haupt, Olaf.Lechtenfeld@itp.uni-hannover.de}
\\[8mm]
\noindent ${}^*${\em
Bogoliubov Laboratory of Theoretical Physics, JINR\\
141980 Dubna, Moscow Region, Russia}\\
{Email: ita, popov@theor.jinr.ru}
\\[8mm]
${}^\times${\em
Centre for Quantum Engineering and Space-Time Research \\
Leibniz Universit\"at Hannover \\
Welfengarten 1, 30167 Hannover, Germany }

\vspace{12mm}

\begin{abstract}
\noindent
Due to their explicit construction, Aloff-Wallach spaces are prominent
in flux compactifications. They carry $G_2$-structures and admit the
$G_2$-instanton equations, which are natural BPS equations for Yang-Mills
instantons on seven-manifolds and extremize a Chern-Simons-type functional.
We consider the Chern-Simons flow between different $G_2$-instantons on
Aloff-Wallach spaces, which is equivalent to Spin(7)-instantons
on a cylinder over them. For a general SU(3)-equivariant gauge connection, 
the generalized instanton equations turn into gradient-flow equations 
on $\C^3\times\R^2$, with a particular cubic superpotential. For the 
simplest member of the Aloff-Wallach family (with 3-Sasakian structure) 
we present an explicit instanton solution of tanh-like shape.
\end{abstract}

\end{center}
\end{titlepage}

\section{Introduction and summary}

Yang-Mills theory in more than four dimensions naturally appears 
in the low-energy limit of superstring theory in the presence of D-branes. 
Also, heterotic strings yield heterotic supergravity, which contains 
supersymmetric Yang-Mills as a subsector~\cite{1}. 
Furthermore, natural Bogomolny-Prasad-Sommerfield-type equations 
for gauge fields in dimension $d{>}4$, introduced in~\cite{2}, 
also appear in superstring compactifications on spacetimes 
$M_{10-d}\times X^d$ as the condition for the survival of at least 
one supersymmetry in the low-energy effective field theory on  
$M_{10-d}$~\cite{1}. 
These first-order Bogomolny-Prasad-Sommerfield-type equations on $X^d$, 
which generalize four-dimensional Yang-Mills anti-self-duality,
were considered e.g.\ in~\cite{3}-\cite{9}, and some of their solutions 
were found in~\cite{10}-\cite{13}.

In string/M-theory compactification, the most interesting dimensions 
seem to be $d=6$, 7 or 8, and the corresponding generalized anti-self-duality 
equations are respectively called the Hermitian-Yang-Mills equations~\cite{4},
the $G_2$-instanton equations~\cite{8,14}, or the Spin(7)-instanton 
equations~\cite{8,15}. Most work on the above-mentioned instanton equations 
has restricted its attention to Riemannian manifolds $X^d$ with holonomy group
SU(3) for $d{=}6$, $G_2$ for $d{=}7$ or Spin(7) for $d{=}8$, i.e.\ to 
{\sl integrable\/} $G$-structures. However, if one is interested in 
string compactification with fluxes~\cite{16}, one should consider 
{\sl non-integrable\/} $G$-structures (weak holonomy groups) instead. 
The torsion of the $G$-structure, which measures the failure to be integrable,
is identified with the three-form field (`flux') of supergravity. 
Flux compactifications have been investigated primarily for type~II strings 
and M-theory, but also in the heterotic theories, albeit to a lesser extent,
despite their long history~\cite{17}. In particular, compactifications 
on Aloff-Wallach spaces~\cite{18,19} $X_{k,l}$ of dimension seven and 
cones $\Ccal(X_{k,l})$ over them were studied e.g.\ in~\cite{20,21,22}. 
The Yang-Mills equations on Spin(7)-manifolds of topology $\R\times X_{k,l}$ 
with cylindrical and conical metric is the subject of the present paper. 
 
For any co-prime pair of integers $(k,l)$, the Aloff-Wallach space $X_{k,l}$ 
is the coset SU(3)/U(1)$_{k,l}$ with U(1)$_{k,l}=\{\textrm{diag} 
(\ep^{\im(k+l)\vp},\ep^{-\im\,k\,\vp},\ep^{-\im\,l\,\vp})\}$~\cite{18,19}.
It carries a $G_2$-structure defined by a torsion three-form~$\psi$ 
with the property that $\diff\psi$ is proportional to the Hodge-dual 
four-form~$\ast\psi$. $G_2$-instantons extremize a Chern-Simons-type action 
functional on~$X_{k,l}$. As an example, we describe the abelian canonical 
connection on a line bundle over $X_{k,l}$. Next, 
we step up to eight dimensions via extending $X_{k,l}$ by a real line $\R$.
Our $G_2$-instantons are the endpoints of a gradient flow along this line,
which is described precisely by the Spin(7)-instanton equations~\cite{8} 
on~$\R\times X_{k,l}$.
The most general SU(3)-equivariant connection on a rank-3 complex vector 
bundle is parametrized by three complex and two real functions on~$\R$. 
The Spin(7)-instanton equations reduce to gradient-flow equations for these
functions, governed by a cubic superpotential~$W$ with global U(1)$\times$U(1)
symmetry. Interestingly, each function obeys a {\sl linear\/} equation
in the background of the others.

In order to be more explicit, we specialize to the case of $k=l=1$.
We fix the metric moduli (up to a freedom of orientation) such that
$X_{1,1}$ is 3-Sasakian and $\Ccal(X_{1,1})$ is hyper-K\"ahler, i.e.\ its
structure group reduces to~Sp(2). We list all critical points and their Hessians
and numerically find an instanton solution whose shape is close to the
tanh function. The corresponding gauge configuration interpolates between
different $G_2$-instantons on~$X_{1,1}$. It is not obvious how to establish
the existence of further instanton solutions. It would be interesting 
to extend and lift Spin(7)-instantons on a cylinder or a cone over 
Aloff-Wallach spaces to classical solutions of heterotic M-theory.

\section{Aloff-Wallach spaces}

\noindent
{\bf Group SU(3).}
Consider the group SU(3) with generators $\{I_a, I_i\}$, $a=1,...,6$, $i=7,8$, satisfying
\begin{equation}\label{2.1}
 [I_a, I_b]= f^c_{ab}I_c + f^i_{ab}I_i\ ,\quad [I_i, I_a]= f^b_{ia}I_b\ ,\quad [I_i, I_j]=0\ ,
\end{equation}
where the structure constants are
\begin{equation}\label{2.2}
\begin{array}{c}
f_{13}^5=f_{42}^5=f_{41}^6=f_{32}^6=-\frac{1}{2\sqrt{3}},\\[6pt]
f_{12}^7=f_{34}^7=\frac{1}{2\sqrt{3}}\ ,\quad
f_{56}^7=-\frac{1}{\sqrt{3}}\and
f_{12}^8=-f_{34}^8=-\frac{1}{2}
\end{array}
\end{equation}
plus those with cyclic permutations of indices in (\ref{2.2}).
The generators (\ref{2.1}) of SU(3) can be chosen in the form
\begin{equation}\label{2.3}
\begin{aligned}
I_1\=\frac{1}{2\sqrt{3}} {\small\begin{pmatrix}
0&0&-1\\0& 0&0\\1&\ 0&0\end{pmatrix}} \ ,\quad
I_3\=\frac{1}{2\sqrt{3}} {\small\begin{pmatrix}
0&1&\ 0\\-1&0&\ 0\\0&0&\ 0\end{pmatrix}} \ ,\quad
I_5\=\frac{1}{2\sqrt{3}} {\small\begin{pmatrix}
0&0&0\\0&0&1\\0&-1&0\end{pmatrix}} \ , \\[8pt]
I_2\=\frac{1}{2\sqrt{3}} {\small\begin{pmatrix}
0&0&\im\\0&0&0\\\im&0&0\end{pmatrix}} \ ,\quad
I_4\=\frac{1}{2\sqrt{3}} {\small\begin{pmatrix}
0&\im&\ 0\\\im&0&0\\0&0&0\end{pmatrix}} \ ,\quad
I_6\=\frac{1}{2\sqrt{3}} {\small\begin{pmatrix}
0&0&0\\0&0&\im\\0&\im&0\end{pmatrix}} \ ,\\[8pt]
I_7\=-\frac{\im}{2\sqrt{3}}\,
{\small\begin{pmatrix}0&0&0\\0&1&0\\0&0&-1\end{pmatrix}}
\und
I_8\=\frac{\im}{6}\,
{\small\begin{pmatrix}2&0&0\\0&-1&0\\0&0&-1\end{pmatrix}}\ ,
\end{aligned}
\end{equation}
corresponding to the anti-fundamental representation.

The basis elements $\{I_a, I_i\}$ of the Lie algebra $su(3)$ can be represented
by left-invariant vector fields $\{E_a, E_i\}$ on the Lie group SU(3), and the dual
basis $\{E^a, E^i\}$ is a set of left-invariant one-forms which obey the Maurer-Cartan
equations
\begin{equation}\label{MC1}
\diff E^a =-f^a_{jb}\, E^j\wedge E^b-\sfrac{1}{2} \,
f^a_{bc}\, E^b\wedge E^c\und \diff E^i = -\sfrac{1}{2} \, f^i_{bc}\,
E^b\wedge E^c \ ,
\end{equation}
where $i,j=7,8$ correspond to the Cartan subalgebra of $su(3)$.

\noindent
{\bf Cosets SU(3)/U(1)$_{k,l}$.}
Let us consider a U(1) subgroup of SU(3) given by matrices of the form
\begin{equation}\label{2.4}
 h=\begin{pmatrix}
\exp(\im(k+l)\vp)&0&0\\0&\exp(-\im k\vp)&0\\0&0&\exp(-\im l\vp)\end{pmatrix}\ ,
\end{equation}
where $k$ and $l$ are relatively prime integers and $0\le\vp\le 2\pi$. Consider the coset space
\begin{equation}\label{2.5}
X_{k,l}=\mbox{SU(3)}/\mbox{U}(1)_{k,l}\ ,
\end{equation}
where U(1)$_{k,l}$ is represented by matrices (\ref{2.4}). For relatively prime integers $k$
and $l$ the coset spaces $X_{k,l}$ are simply connected manifolds called Aloff-Wallach spaces~\cite{18,19}.

The space SU(3)/U(1)$=:G/H$ consists of left cosets $gH$, $g\in G$, and the natural projection
$g\mapsto gH$ is denoted by
\begin{equation}\label{2.6}
\pi :\quad\mbox{SU(3)}\longrightarrow X_{k,l}
\end{equation}
with fibres U(1)$_{k,l}$. Over a contractible open subset $U$ of $X_{k,l}$, one can choose a map
$L:\ U\to $SU(3) such that $\pi\circ L=\mbox{Id}_U$, i.e. $L$ is a local section of the principal bundle
(\ref{2.6}). The pull-backs of $\{E^a, E^i\}$ by $L$ from SU(3) to $X_{k,l}$ are denoted by $\{e^a, e^i\}$
which satisfy the same Maurer-Cartan equations
\begin{equation}\label{MC}
\diff e^a =-f^a_{ib}\, e^i\wedge e^b-\sfrac{1}{2} \,
f^a_{bc}\, e^b\wedge e^c\und \diff e^i = -\sfrac{1}{2} \, f^i_{bc}\,
e^b\wedge e^c
\end{equation}
as $\{E^a, E^i\}$. Note that since all objects we consider will be invariant under some action of
SU(3), it will suffice to do calculations just over the subset $U$.

If we denote by $\{e^{\ah}\}$, $\ah =1,...,7$, an orthonormal coframe on $U\subset X_{k,l}$ (basis for
$T^*(X_{k,l})$ over $U$) then
\begin{equation}\label{2.9}
e^{\ah}=e^a \for a=1,...6\und e^{\hat 7}=\sfrac{1}{\De}\,(k+l)\,e^7 - \sfrac{2\g}{\De}\,(k-l)\,e^8
\end{equation}
with $\{e^a, e^i\}$ obeying the Maurer-Cartan equations (\ref{MC}) and
\begin{equation}\label{2.10a}
e^{\hat 8}= \sfrac{1}{2\g\De}\,(k-l)\,e^7 + \sfrac{1}{\De}\,(k+l)\,e^8
\end{equation}
is a canonical connection one-form in the bundle (\ref{2.6}). Here
\begin{equation}\label{2.10b}
\g :=\sfrac{1}{2\sqrt{3}}\und \De^2:=2(k^2+l^2)\ .
\end{equation}
Then as generators of SU(3) we have
\begin{equation}\label{Ih}
I_{\ah}=I_a\ ,\quad I_{\hat 7}= \De^{-1}\left((k{+}l)I_7-\sqrt{3}(k{-}l)I_8\right)
\and I_{\hat 8}= \De^{-1}\left(\sfrac{1}{\sqrt{3}}(k{-}l)I_7+(k{+}l)I_8\right),
\end{equation}
so that
\begin{equation}\label{2.11a}
 e^aI_a + e^iI_i=e^{\ah}I_{\ah} + e^{\hat 8}I_{\hat 8}
\end{equation}
and $I_{\hat 8}$ is the generator of the group U(1)$_{k,l}$ given by (\ref{2.4}).

Let us now rescale matrices (\ref{Ih}) as
$$
\It_1=\g^{-1}\s_1 I_1\,,\  \It_2=\g^{-1}\s_1 I_2\,,\quad \It_3=\g^{-1}\s_2 I_3 \,,\
\It_4=\g^{-1}\s_2 I_4\, ,
$$
\begin{equation}\label{It}
\quad \It_5=\g^{-1}\s_3 I_5\,,\ \It_6=\g^{-1}\s_3 I_6 \ ,\quad
\It_7= (\g\mu)^{-1}I_{\hat 7}\,,\
\It_8= (\g\mu)^{-1}I_{\hat 8}\ .
\end{equation}
such that
\begin{equation}\label{2.11}
 e^aI_a + e^iI_i=\et^{\ah}\It_{\ah} + \et^{8}\It_{8}\ ,
\end{equation}
and therefore the rescaled coframe fields $\{\et^{\ah}\}$ and the rescaled connection one-form
$\et^{8}$ have the form
$$
\et^1=\g\s_1^{-1}e^1,\ \et^2=\g\s_1^{-1}e^2,\quad \et^3=\g\s_2^{-1}e^3,\ \et^4=\g\s_2^{-1}e^4,
$$
\begin{equation}\label{et}
\et^5=\g\s_3^{-1}e^5,\ \et^6=\g\s_3^{-1}e^6,\quad
 \et^7=\g\mu\,e^{\hat 7},\
\et^8=\g\mu\,e^{\hat 8}\ .
\end{equation}
Here we introduced real parameters
\begin{equation}\label{sm}
\s_1\,,\ \s_2\,,\  \s_3\,,\ \mu\in\R\ .
\end{equation}

As a metric on $X_{k,l}$ we take
\begin{equation}\label{metric1}
 \diff\tilde{s}^2_7= \delta_{\ah\bh}\,\et^{\ah}\et^{\bh}\ .
\end{equation}
One can show that for any given relatively prime integers $k,l$ one can choose parameters $\s_{\a}$ and
$\mu$ ($\a =1,2,3$) such that the metric (\ref{metric1}) will be Einstein for a connection
with a torsion 3-form
\begin{equation}\label{torsion}
 \psi=\frac{1}{3!}\psi_{\ah\bh\hat c}\et^{\ah}\wedge\et^{\bh}\wedge\et^{\hat c}
\end{equation}
having the following non-vanishing components:
\begin{equation}\label{torsioncomp}
 \psi_{135}= \psi_{425}= \psi_{416}= \psi_{326}= \psi_{127}= \psi_{347}=\psi_{567}=1 \ .
\end{equation}
Furthermore, this connection has the holonomy group $G_2$ and the 3-form (\ref{torsion}) defines
a $G_2$-structure on $X_{k,l}$~\cite{18,19}. For more details on geometry of Aloff-Wallach spaces
see e.g.~\cite{18}-\cite{21}.

\bigskip

\noindent
{\bf Complex basis on $T^*(X_{k,l})$.} Note that $X_{1,1}$ can be fibred over the homogeneous manifold
$\F_3=$SU(3)$/$U(1)$\times$U(1) with fibres
\begin{equation}\label{U1}
\mbox{U(1)}^{\bot}=\exp(\a\,\It_7)
\end{equation}
parametrized by an angle $0\le\a\le 2\pi$. So, for $k=l=1$ we have a projection
\begin{equation}\label{XF}
X_{1,1}\longrightarrow \F_3\ ,
\end{equation}
whose fibres U(1)$^\bot$ are orthogonal complements of U(1)=U(1)$_{1,1}$ from (\ref{2.4})-(\ref{2.6})
in the torus $T^2\cong$ U(1)$\times$U(1) (the Cartan subgroup of SU(3)). This case is very special since
$X_{1,1}$ is an Einstein-Sasaki manifold and therefore the cone $\Ccal (X_{1,1})$ with the metric
\begin{equation}\label{metric8}
\diff s^2_8=\diff r^2 + r^2\diff s^2_{X_{1,1}}
\end{equation}
is a Calabi-Yau 4-conifold with the holonomy group\footnote{Recall that the cone $\Ccal(X_{k,l})$ over 
the general Aloff-Wallach space $X_{k,l}$ has the holonomy group Spin(7).} SU(4)$\,\subset\,$Spin(7). Furthermore, on $X_{1,1}$ there exists a metric such that $X_{1,1}$ becomes a 3-Sasakian manifold with a hyper-K\"ahler structure Sp(2) on the cone  $\Ccal (X_{1,1})$.

Recall that $\F_3$ is fibred over the projective plane $\C P^2\cong\ $SU(3)$/$U(2),
$\F_3 \longrightarrow\C P^2$, and the same is true for $X_{k,l}$ with any $k$ and $l$. One can show that fibres of the projection $X_{k,l}\longrightarrow \C P^2$ are lens spaces $S^3/\Z_p$ with $p=|k+l|$. For clarity, let 
us combine all the above fibrations into one diagram
\bigskip
\begin{equation}\label{CD}
\begin{CD}
\mbox{SU(3)}@>{\rm U}(1)\times {\rm U}(1)>>\F_3\\
@VV{\rm U}(1)_{k,l}V              @VV\C P^1V\\
X_{k,l}     @>S^3/\Z_p>>      \C P^2
\end{CD}
\end{equation}
where $X_{k,l}$ can also be fibred over $\F_3$ if $k=l=1$.

Note that $S^3/\Z_p$ is an $S^1$-fibre bundle over $\C P^1$ and one can consider complex forms which 
span $\C P^2$ and $\C P^1$ in $X_{k,l}$ as seen from (\ref{CD}). Namely, let us introduce complex 
one-forms\footnote{Here, $\Th^{1,2}$ span the $\C P^2$ base in (\ref{CD}) and $\Th^{3}$ spans the $\C P^1\hra S^3/\Z_p$, and the choice of the sign in $\Th^{3}$ is such that an associated almost complex
structure on a six-dimensional subbundle of the tangent bundle, defined by $\Th^{\a}, \a=1,2,3$, 
will be integrable. For $\Th^{3}=\et^5 + \im \et^6$ it will be never integrable. For $k{=}l{=}1$ our choice corresponds to a K\"ahler structure on $\F_3$ and $\Th^{3}=\et^5 + \im \et^6$ corresponds to a nearly K\"ahler structure on $\F_3$~\cite{9,23}.}
\begin{equation}\label{Th}
\begin{matrix}
\Th^{1}:=\et^{1}+\im\et^{2}\, ,\  \Th^{2}:=\et^{3}+\im\et^{4}\, ,\  \Th^{3}=-\et^5 + \im \et^6\\[6pt]
\Th^{\1}:=\et^{1}-\im\et^{2}\, ,\  \Th^{\2}:=\et^{3}-\im\et^{4}\, ,\  \Th^{\3}=-\et^5 - \im \et^6
\end{matrix}
\end{equation}
plus real $\et^7, \et^8$ and matrices
\begin{equation}\label{Ita}
\begin{matrix}
 \It^-_1 :=\sfrac12(\It_{1}-\im\It_{2})\, ,\   \It^-_2 :=\sfrac12(\It_{3}-\im\It_{4})\, ,\   
\It^-_3 :=\sfrac12(-\It_{5}-\im\It_{6})\, ,\\[6pt]
\It^+_{\1} :=\sfrac12(\It_{1}+\im\It_{2})\, ,\   \It^+_{\2} :=\sfrac12(\It_{3}+\im\It_{4})\, ,\   
\It^+_{\3} :=\sfrac12(-\It_{5}+\im\It_{6})\and
 -\im\It_{7}\, ,\ -\im\It_{8}\ ,
\end{matrix}
\end{equation}
which form a basis of the complex Lie algebra $sl(3,\C)=su(3)\otimes\C$.  Their explicit form is
\bea\nonumber
&\It_1^-=\s_1\,\small{
\begin{pmatrix}0&0&0\\0&0&0\\1&0&0\end{pmatrix} }\ ,&
\It_{\1}^+=\s_1\,
\small{\begin{pmatrix}0&0&-1\\0&0&0\\0&0&0\end{pmatrix}}\ , \\[4pt]
\nonumber
&\It_2^-=\s_2\,
\small{\begin{pmatrix}0&1&0\\0&0&0\\0&0&0\end{pmatrix}}\ ,&
\It_{\2}^+=\s_2\,
\small{\begin{pmatrix}0&0&0\\-1&0&0\\0&0&0\end{pmatrix}}\ , \\[4pt]
\label{III}
&\It_3^-=\s_3
\small{\begin{pmatrix}0&0&0\\0&0&0\\0&1&0\end{pmatrix}}  \ ,&
\It_{\3}^+=\s_3\,
\small{\begin{pmatrix}0&0&0\\0&0&-1\\0&0&0\end{pmatrix}}\ , \\[4pt]
&-\im\,\It_7=\frac{2}{\mu\Delta}\,
\small{\begin{pmatrix}l-k&0&0\\0&-l&0\\0&0&k\end{pmatrix}} \and&
-\im\,\It_8=\frac{2}{\mu\Delta\sqrt{3}}\,
\small{\begin{pmatrix}k+l&0&0\\0&-k&0\\0&0&-l\end{pmatrix}}\ .
\nonumber\eea

We have the commutation relations
\begin{equation}\label{II}
\begin{matrix}
[-\im\It_j, \It^-_\a ]=\Ct^\b_{j\a}\It_\b^-\ , \quad [-\im\It_j, \It^+_{\ab} ]=\Ct^{\bb}_{j\ab}\It^+_{\bb}\ , \\[3mm]
[\It^-_\a, \It^-_\b ]=\Ct^{\g}_{\a\b}\It^-_{\g}\ , \quad
[\It^+_{\ab},\It^+_{\bb}]=\Ct^{\bar\g}_{\ab\bb}\It^+_{\bar\g}\ , \\[3mm]
 [\It^-_\a, \It^+_{\bb}]=\Ct^j_{\a\bb}(-\im\,\It_j)+\Ct^{\g}_{\a\bb}\It^-_{\g} +\Ct^{\bar\g}_{\a\bb}\It^+_{\bar\g}
\end{matrix}
\end{equation}
with
\begin{equation}\label{Ct}
\begin{matrix}
\Ct^{1}_{3\2}=-\frac{\s_2\,\s_3}{\s_1}=-\Ct^{\1}_{2\3}\ ,\quad
\Ct^{2}_{3\1}=\frac{\s_3\,\s_1}{\s_2}= -\Ct^{\2}_{1\3}\ ,\quad
\Ct^{3}_{12}=\frac{\s_1\,\s_2}{\s_3}= \Ct^{\3}_{\1\2}\ ,\\[5mm]
\Ct^1_{71}=\frac{2}{\mu\De}(2k-l)=-\Ct^{\1}_{7\1}\ ,\quad
\Ct^2_{72}=\frac{2}{\mu\De}(2l-k)=-\Ct^{\2}_{7\2}\ ,\quad
\Ct^3_{73}=\frac{2}{\mu\De}(k+l)=-\Ct^{\3}_{7\3}\ ,\\[5mm]
\Ct^1_{81}{=}{-}\frac{2}{\sqrt{3}\mu\De}(k{+}2l){=}{-}\Ct^{\1}_{8\1}\ ,\quad
\Ct^2_{82}{=}\frac{2}{\sqrt{3}\mu\De}(2k{+}l){=}{-}\Ct^{\2}_{8\2}\ ,\quad
\Ct^3_{83}{=}\frac{2}{\sqrt{3}\mu\De}(k{-}l){=}{-}\Ct^{\3}_{8\3}\ ,\\[5mm]
\Ct^7_{1\1}=-\frac{\mu\s_1^2}{\De}k\ ,
\qquad
\Ct^7_{2\2}=-\frac{\mu\s_2^2}{\De}l \ ,
\qquad
\Ct^7_{3\3}=-\frac{\mu\s_3^2}{\De}(k+l) \ ,\\[5mm]
\Ct^8_{1\1}=\frac{\sqrt{3}\mu\s_1^2}{\De}l\ ,
\qquad
\Ct^8_{2\2}=-\frac{\sqrt{3}\mu\s_2^2}{\De}k \ ,
\qquad
\Ct^8_{3\3}=\frac{\sqrt{3}\mu\s_3^2}{\De}(l-k) \ .
\end{matrix}
\end{equation}
Note that standard undeformed structure constants correspond to $k{=}l{=}1$, $\mu{=}\g^{-1}$, $\s_1{=}\s_2{=}\g$, $\s_3{=}\sfrac{\g}{\sqrt{2}}$ and they are given by
\begin{equation}\label{C}
\begin{matrix}
C^{1}_{3\2}=-\sfrac{1}{\sqrt{2}}\g =-C^{\1}_{2\3}\ ,\quad C^{2}_{3\1}=\sfrac{1}{\sqrt{2}}\g =-C^{\2}_{1\3}\ ,
\quad C^{3}_{12}=\sqrt{2}\g=C^{\3}_{\1\2} ,\\[4mm]
C^{1}_{71}=C^{2}_{72}=-C^{\1}_{7\1}=-C^{\2}_{7\2}=\g\ ,
\qquad C^{3}_{73}=2\g =-C^{\3}_{7\3}\ ,\\[4mm]
C^{1}_{81}=-\frac{1}{2}=-C^{\1}_{8\1}\ ,
\qquad C^{2}_{82}=\frac{1}{2}=-C^{\2}_{8\2}\ ,\\[4mm]
C^{7}_{1\1}=C^{7}_{2\2}=C^{7}_{3\3}=-\sfrac{1}{2}\g\ ,
\qquad C^{8}_{1\1}=\frac{1}{4}=-C^{8}_{2\2}\ .
\end{matrix}
\end{equation}

In the new basis the Maurer-Cartan equations (\ref{MC}) become
\begin{equation}\label{MCTh}
\begin{matrix}
\diff \Th^{\a} =-\im \Ct^{\a}_{j\b}\, \et^j\wedge \Th^{\b}-\sfrac{1}{2} \,
\Ct^{\a}_{\b\g }\, \Th^{\b}\wedge \Th^{\g}-\Ct^{\a}_{\b\bar\g }\, \Th^{\b}\wedge \Th^{\bar\g}\ ,\\[3mm]
\diff \Th^{\ab} =-\im \Ct^{\ab}_{j\bb}\, \et^j\wedge \Th^{\bb}-\sfrac{1}{2} \,
\Ct^{\ab}_{\bb\bar\g }\, \Th^{\bb}\wedge \Th^{\bar\g} - \Ct^{\ab}_{\b\bar\g }\, \Th^{\b}\wedge \Th^{\bar\g}\ ,\\[3mm]
\diff \et^j = \im \, \Ct^j_{\b\bar\g}\,\Th^{\b}\wedge \Th^{\bar\g} \ ,
\end{matrix}
\end{equation}
where we have used the structure constants from (\ref{Ct}).
The metric on $X_{k,l}$ in terms of $\Th^{\a}$ and $\et^7$ is
\begin{equation}\label{metric7}
 \diff\tilde{s}^2_7= \delta_{\a\bb}\,\Th^{\a}\Th^{\bb} + (\et^7)^2\ ,
\end{equation}
i.e. we have
\begin{equation}\label{metr}
g_{\a\bb}=\sfrac12\,\delta_{\a\bb} \und g_{77}=1\ .
\end{equation}

\bigskip

\noindent
{\bf Coset space $X_{1,1}$.} It is known (see e.g.~\cite{19}) that for $k{=}l{=}1$ the Aloff-Wallach 
space admits a metric such that the cone $\Ccal (X_{1,1})$ over it admits metrics with the holonomy group SU(4)$\,\subset\,$Spin(7) (Calabi-Yau 4-fold) and Sp(2)$\,\subset\,$SU(4)$\,\subset\,$Spin(7) (hyper-K\"ahler 4-fold). This means that in the Calabi-Yau case on $\Ccal (X_{1,1})$ there exists a closed (1,1)-form 
$\omega^{1,1}$ (K\"ahler form) and in the hyper-K\"ahler case on $\Ccal (X_{1,1})$ there exist three K\"ahler
forms:
\begin{equation}\label{3o}
 \om_3=\om^{1,1}\ ,\quad \om_1\and\om_2\ ,
\end{equation}
i.e. besides the closed form $\om^{1,1}$ we also have a closed (2,0)-form $\om^{2,0}:=\om_1+\im\,\om_2$.

For the general metric (\ref{metric8}) on $\Ccal (X_{1,1})$, one can introduce the (1,1) form as
\begin{equation}\label{o11}
\om^{1,1}:=\sfrac{\im}{2}\,r^2\,(\delta_{\a\ab}\Th^{\a}\wedge\Th^{\ab}+\Th^{4}\wedge\Th^{\4})\ ,
\end{equation}
where
\begin{equation}\label{o4}
\Th^{4}:=\frac{\diff r}{r}-\im\,\et^7\und  \Th^{\4}:=\frac{\diff r}{r}+\im\,\et^7\ .
\end{equation}
We obtain
$$
-2\,\im\,\diff \om^{1,1}= (2-\mu\s_1^2)\Th^{1\1}\wedge r\diff r  
+ (2-\mu\s_2^2)\Th^{2\2}\wedge r\diff r + (1-\mu\s_3^2)\Th^{3\3}\wedge 2r\diff r 
$$
\begin{equation}\label{do11}
+\ r^2\left(\frac{\s_2\s_3}{\s_1}+\frac{\s_3\s_1}{\s_2} - 
\frac{\s_1\s_2}{\s_3}\right)(\Th^{12\3}-\Th^{\1\2 3})\  ,
\end{equation}
where $\Th^{\a\bb}:=\Th^{\a}\wedge\Th^{\bb}$ etc.
From (\ref{do11}) it follows that $\om^{1,1}$ is closed if 
\begin{equation}\label{sss}
 \s_1^2 = \s_2^2 = 2 \s_3^2 = 2\a^2 \and \mu=\frac{1}{\a^{2}}
\end{equation}
for any real number $\a$.

\bigskip

\noindent
{\bf SU(4)- and Sp(2)-holonomy on $\Ccal (X_{1,1})$.}
Note that the closure of the form $\omega^{1,1}$ means that the holonomy group of the cone $\Ccal (X_{1,1})$
reduces to the group U(4) (K\"ahler structure). For having on $\Ccal (X_{1,1})$ a Calabi-Yau structure
(SU(4)-holonomy) one should impose an additional condition of closure of the (4,0)-form
\begin{equation}\label{O40}
\Omega^{4,0}:= r^4\Th^1\wedge\Th^2\wedge\Th^3\wedge\Th^4\ .
\end{equation}
By differentiating (\ref{O40}), from the condition $\diff\Omega^{4,0}=0$ one obtains
\begin{equation}\label{aa}
\a=+1\qquad\mbox{or}\qquad\a=-1
\end{equation}
that fixes a Calabi-Yau metric on $\Ccal (X_{1,1})$. Both $\a$ from (\ref{aa}) correspond to the same metric
on $X_{1,1}$ and the choice of different sign of $\a$ corresponds to the choice of different orientation
on $X_{1,1}$.

Now we want to check whether this metric allows further reduction of the structure group SU(4)
to the group Sp(2)$\,\subset\,$SU(4)$\,\subset\,$Spin(7), i.e. allows an introduction of a hyper-K\"ahler structure on $\Ccal (X_{1,1})$. On the Calabi-Yau space $\Ccal (X_{1,1})$, we consider the (2,0)-form
\begin{equation}\label{o20}
 \om^{2,0}= r^2 (\Th^1\wedge\Th^2+\b\,\Th^3\wedge\Th^4)\ ,
\end{equation}
where $\b$ is a complex number. Then from the equation $\diff\om^{2,0}=0$ we obtain
\begin{equation}\label{ab}
\b=\a=\pm\,1\ .
\end{equation}
Therefore, the metric with $\a=\pm 1$ from (\ref{aa}) allows a hyper-K\"ahler structure\footnote{Comparing with the standard expression for the symplectic form in Darboux coordinates, the careful reader might notice an unconventional relative sign appearing in~(\ref{o20}) for the choice $\a=-1$. To arrive at the standard expression $\om^{2,0}= \Th^1\wedge\Th^2+\Th^3\wedge\Th^4$, which is unique up to an overall rescaling, one needs to absorb the sign by replacing $\Th^4$ with minus itself in the definition~(\ref{o4}). This has no further consequences except for an irrelevant overall sign-flip in~(\ref{O40}) corresponding to the change of orientation.} on the cone $\Ccal (X_{1,1})$ and a 3-Sasakian structure on $X_{1,1}$.

\section{Spin(7)-instantons}

\noindent
{\bf $G_2$-instantons and gradient flows.}
Consider the Chern-Simons type functional on $X_{k,l}$,
$$
S=-\sfrac14\int\limits_{X_{k,l}}\tr\,(\Fcal\wedge\Fcal )\wedge\psi=
-\sfrac14\int\limits_{X_{k,l}}\tr(\Acal\wedge\diff\Acal +
\sfrac23\Acal\wedge\Acal\wedge\Acal )\wedge\diff\psi 
$$
\begin{equation}\label{3.1}
-\ \sfrac14\int\limits_{X_{k,l}}\diff (\tr(\Acal\wedge\diff\Acal +\sfrac23\Acal\wedge\Acal\wedge\Acal )\wedge\psi )\ ,
\end{equation}
where $\Acal$ is a connection on a rank-3 complex vector bundle over $X_{k,l}$ 
(we will specialise to the gauge group SU$(3)$ in a moment) 
and $\Fcal =\diff\Acal +
\Acal\wedge\Acal$ is its curvature.
For the variation of (\ref{3.1}) we have
\begin{equation}\label{3.2}
\left(\frac{\delta S}{\delta\Acal}\right )_{\ah}=\sfrac12 \ast (\diff\psi\wedge\Fcal )_{\ah}= 
\sfrac12\,\b\,\psi_{\ah\bh\ch}\Fcal_{\bh\ch}\ ,
\end{equation}
where $\ast$ is the Hodge operator and $\b$ is some coefficient
which can be calculated. Here, we used the fact that 
$\diff\psi\sim\ast\psi\ \Rightarrow\ \ast\diff\psi\sim\psi$ on $X_{k,l}$. 
Therefore, the equations of motion are
\begin{equation}\label{3.3}
\diff\psi\wedge\Fcal =0\quad\Leftrightarrow\quad\psi\lrc\Fcal =0\quad\Leftrightarrow\quad
\psi_{\ah\bh\ch}\Fcal_{\bh\ch}=0\ .
\end{equation}
Note that (\ref{3.3}) are exactly $G_2$-instanton equations on $X_{k,l}$.
Now we can define the Chern-Simons gradient flow equations
\begin{equation}\label{3.4}
\dot\Acal_{\ah}=\b^{-1}\left(\frac{\delta S}{\delta\Acal}\right )_{\ah}
=\sfrac12\,\psi_{\ah\bh\ch}\Fcal_{\bh\ch}\ ,
\end{equation}
whose stable points $\dot\Acal:=\frac{\diff\Acal}{\diff\tau}=0$ are $G_2$-instantons on $X_{k,l}$.

\bigskip

\noindent
{\bf Spin(7)-instanton equations on $\R\times X_{k,l}$.}
On the one hand, (\ref{3.4}) are the flow equations. 
On the other hand, they are exactly the Spin(7)-instanton equations
\begin{equation}\label{3.5}
\Fcal_{0\ah}=\sfrac12\,\psi_{\ah\bh\ch}\Fcal_{\bh\ch}
\end{equation}
on the space $\R\times X_{k,l}$, $\tau\in\R$, in the gauge $\Acal_{\tau}\equiv\Acal_0=0$, where
$\tau\equiv x^0$, $\diff\tau =\et^0$.

So, let us consider $\Acal , \Fcal\in su(3)$,
and the equations (\ref{3.5}) on the space $\R\times X_{k,l}$.
The SU(3)-equivariant ansatz for $\Acal$ is
\begin{equation}\label{3.6}
\Acal = X_{\ah}\et^{\ah} + \It_8\et^8 =Y_{\a}\Th^{\a} + Y_{\ab}\Th^{\ab} + X_7\et^7 +\It_8\et^8\ ,
\quad \Acal_0=0\ ,
\end{equation}
with the following restrictions which guarantee the SU(3)-equivariance:
\begin{equation}\label{3.7}
[-\im\,\It_8, Y_{\a}]=\Ct^{\b}_{8\a}Y_{\b}\ ,\quad [-\im\,\It_8, Y_{\ab}]=\Ct^{\bb}_{8\ab}Y_{\bb}\ ,\quad
[\It_8, X_7]=0\ .
\end{equation}
Here
\begin{equation}\label{3.6a}
Y_1 :=\sfrac{1}{2}\, (X_1-\im \, X_2),\quad Y_2 :=\sfrac{1}{2}\, (X_3-\im \, X_4),\quad
Y_3 :=\sfrac{1}{2}\, (-X_5-\im \, X_6)\and Y_{\ab}=-Y_\a^\+
\end{equation}
are some $3{\times}3$ complex matrices depending on $\tau\in\R$, $\a , \b =1,2,3$.

For (\ref{3.6}) and (\ref{3.7}) we have
\bea
\Fcal &=&\dot Y_{\a}\,\et^0\wedge \Th^{\a}+ \dot Y_{\ab}\,\et^0\wedge \Th^{\ab} +\dot X_7\,\et^0\wedge\et^7 +
\sfrac{1}{2} \, \big([Y_\a, Y_\b]-\Ct_{\a\b}^{\g}\, Y_{\g}\big)\, \Th^\a\wedge \Th^\b 
\nonumber \\
&+&\big([Y_{\a}, Y_{\bb}]-\Ct_{\a\bb}^{\g}\,Y_{\g}-\Ct_{\a\bb}^{\bar\g}\,Y_{\bar\g}+\im\,\Ct_{\a\bb}^7\, X_7 +\im\,\Ct_{\a\bb}^8\, \It_8\big)\, \Th^\a\wedge\Th^{\bb}\label{3.8a}\\
&+&\sfrac{1}{2}\big([Y_{\ab}, Y_{\bb}]{-}\Ct_{\ab\bb}^{\bar\g}Y_{\bar\g}\big) \Th^{\ab}{\wedge}\Th^{\bb}
+ \big([Y_{\a}, X_{7}]+\im\,\Ct^{\b}_{7\a}Y_{\b}\big)\Th^\a{\wedge}\et^7 +\big([Y_{\ab}, X_{7}]+
\im\,\Ct^{\bb}_{7\ab}Y_{\bb}\big)\Th^{\ab}{\wedge}\et^7\,,
\nonumber\eea
where $\dot Y_\a:=\diff Y_\a/\diff\tau$ etc.
We get
\begin{equation}\label{3.8}
\begin{matrix}
\Fcal_{0\a}=\dot Y_{\a}\ ,\quad\Fcal_{0\ab}=\dot Y_{\ab}\ ,\quad\Fcal_{07}=\dot X_7\ ,\\[2mm]
\Fcal_{\a\b}=[Y_{\a}, Y_{\b}] - \Ct^{\g}_{\a\b}Y_{\g}\ ,\quad
\Fcal_{\ab\bb}=[Y_{\ab}, Y_{\bb}] - \Ct^{\bar\g}_{\ab\bb}Y_{\bar\g}\ ,\\[2mm]
\Fcal_{\a\bb}=[Y_{\a}, Y_{\bb}]- \Ct^{\g}_{\a\bb}Y_{\g}- \Ct^{\bar\g}_{\a\bb}Y_{\bar\g} +
\im\,\Ct^{7}_{\a\bb}X_{7}+\im \Ct^{8}_{\a\bb}\It_{8}\ ,\\[2mm]
\Fcal_{\a 7}=[Y_{\a}, X_{7}] +\im \Ct^{\b}_{7\a}Y_{\b}\ ,\quad
\Fcal_{\ab 7}=[Y_{\ab}, X_7] + \im\Ct^{\bb}_{7\ab}Y_{\bb}\ .
\end{matrix}
\end{equation}

\bigskip

\noindent
{\bf Reduction to matrix equations.} Note that
\begin{equation}\label{3.9}
\psi = \sfrac{1}{3!}\,\psi_{\ah\bh\ch}\ \et^{\ah\bh\ch}= \psi_{12\3}\Th^{12\3}+
\psi_{\1\2 3}\Th^{\1\2 3}+\sfrac{1}{2!}\,\psi_{7\a\bb}\ \et^7\wedge\Th^{\a\bb}\ ,
\end{equation}
and therefore
\begin{equation}\label{3.10}
\psi_{12\3}=\psi_{\1\2 3}=-\sfrac12 ,\quad  \psi_{71\1}=\psi_{72\2}=\sfrac{\im}{2}\and
\psi_{73\3}=-\sfrac{\im}{2}\ .
\end{equation}
Thus, from (\ref{3.5}) we have
\begin{equation}\label{3.11}
2\Fcal_{0\a}= -2\psi_{\a\b\bar\g}\,\Fcal_{\g\bb} + \psi_{\a\bb\bar\g}\,\Fcal_{\b\g}+
2\psi_{\a\bb 7}\,\Fcal_{\b 7}\ ,
\end{equation}
\begin{equation}\label{3.12}
2\Fcal_{0\ab}= 2\psi_{\ab\bb\g}\,\Fcal_{\b\bar\g} + \psi_{\bar\a\b\g}\,\Fcal_{\bb\bar\g}+
2\psi_{\ab\b 7}\,\Fcal_{\bb 7}\ ,
\end{equation}
\begin{equation}\label{3.13}
2\Fcal_{0 7}= \psi_{7\ah\bh}\Fcal_{\ah\bh}=-2\psi_{7\a\ab}\Fcal_{\a\ab}\ .
\end{equation}
Substituting (\ref{3.8})-(\ref{3.10}) into (\ref{3.11})-(\ref{3.13}), we obtain the following matrix equations
\begin{equation}\label{3.14}
 2\dot Y_{\1}= (\Ct^{\1}_{2\3}+\Ct^{\1}_{7\1})Y_{\1} +[\im\,X_7, Y_{\1}]- [Y_2, Y_{\3}] \ ,
\end{equation}
\begin{equation}\label{3.15}
 2\dot Y_{\2}= (\Ct^{\2}_{\3 1}+\Ct^{\2}_{7\2})Y_{\2}+[\im\,X_7, Y_{\2}] - [Y_{\3}, Y_1] \ ,
\end{equation}
\begin{equation}\label{3.16}
 2\dot Y_{\3}= (\Ct^{\3}_{\1\2}-\Ct^{\3}_{7\3})Y_{\3} -[\im\,X_7, Y_{\3}] - [Y_{\1}, Y_{\2}] \ ,
\end{equation}
\begin{equation}\label{3.17}
 2\dot X_7=\Ct^8\It_8 + \Ct^7 X_7  - \im\,\left([Y_{1}, Y_{\1}]+[Y_{2}, Y_{\2}]-[Y_{3}, Y_{\3}]\right )\ ,
\end{equation}
where 
\begin{equation}\label{3.17a}
\Ct^8:=\Ct^8_{1\1}+\Ct^8_{2\2}-\Ct^8_{3\3}\and \Ct^7:=\Ct^7_{1\1}+\Ct^7_{2\2}-\Ct^7_{3\3}\ .
\end{equation}
All structure constants in (\ref{3.14})-(\ref{3.17a}) can be taken from (\ref{Ct}). The above matrix equations can be written concisely by means of a ``superpotential''~$W$ via
\begin{equation}
	\dot{Y}_{\bar{\alpha}} = \delta_{\bar{\alpha}\beta} \frac{\partial W}{\partial Y_\beta} \; , \qquad\qquad 
	\dot{X}_7 = \frac{\partial W}{\partial X_7} \; .
\end{equation}
The explicit form of the superpotential $W(Y_1,Y_2,Y_3,Y_{\bar{1}},Y_{\bar{2}},Y_{\bar{3}},X_7)$,
\begin{multline}
	W = \frac{1}{2} \tr\left\{\vphantom{\frac{1}{2}} (\tilde{C}^{\bar{1}}_{2\bar{3}} + \tilde{C}^{\bar{1}}_{7\bar{1}}) Y_1 Y_{\bar{1}} + (\tilde{C}^{\bar{2}}_{\bar{3} 1} + \tilde{C}^{\bar{2}}_{7\bar{2}}) Y_2 Y_{\bar{2}} + (\tilde{C}^{\bar{3}}_{\bar{1}\bar{2}} - \tilde{C}^{\bar{3}}_{7\bar{3}}) Y_3 Y_{\bar{3}} \right.\\\left. - \left[Y_{\bar{2}}, Y_3 \right] Y_{\bar{1}} - \left[Y_{\bar{3}}, Y_1 \right] Y_2 + i \left( \left[X_7, Y_{\bar{1}} \right] Y_1 + \left[X_7, Y_{\bar{2}} \right] Y_2 - \left[X_7, Y_{\bar{3}} \right] Y_3 \right) \right.\\\left. + \tilde{C}^8 \tilde{I}_8 X_7 + \frac{1}{2} \tilde{C}^7 (X_7)^2 \right\} \; ,
\end{multline}
follows by inspection of (\ref{3.14})-(\ref{3.17a}). It can also be obtained directly by inserting the ansatz~(\ref{3.6}) into the Chern-Simons type action~(\ref{3.1}).

\bigskip

\noindent
{\bf Reduction to equations on scalar fields of $\tau$.}
The SU(3)-equivariance conditions (\ref{3.7}) are solved by
\begin{equation}\label{3.18}
\begin{matrix}
Y_{\1}=\bar\phi^{\1}\It_{\1}\ ,\quad  Y_{\2}=\bar\phi^{\2}\It_{\2}\ ,\quad Y_{\3}=\bar\phi^{\3}\It_{\3}\ ,\\[4mm]
Y_{1}=\phi^1\It_{1}\ ,\quad  Y_{2}=\phi^2\It_{2}\ ,
\quad Y_{3}=\phi^3\It_{3}\ ,\\[4mm]
X_7=\chi^7\It_7 + \chi^8\It_8\ ,
\end{matrix}
\end{equation}
where $\phi^{\a}$ ($\a =1,2,3$) are complex scalar fields depending on $\tau$ and $\chi^i$ ($i=7,8$)
are real scalar fields of $\tau$.

Substituting (\ref{3.18}) into (\ref{3.14})-(\ref{3.17}), we obtain
\begin{equation}\label{3.19}
\begin{matrix}
2\dot\phi^1=(\Ct^{\1}_{2\3}+\Ct^{\1}_{7\1}-\chi^7\Ct^{\1}_{7\1}-\chi^8\Ct^{\1}_{8\1})\phi^1-
\Ct^{\1}_{2\3}\bar\phi^{\2}\phi^{3}\ ,\\[4mm]
2\dot\phi^2=(\Ct^{\2}_{\31}+\Ct^{\2}_{7\2}-\chi^7\Ct^{\2}_{7\2}-\chi^8\Ct^{\2}_{8\2})\phi^2-
\Ct^{\2}_{\31}\bar\phi^{\1}\phi^{3}\ ,\\[4mm]
2\dot\phi^3=(\Ct^{\3}_{\1\2}-\Ct^{\3}_{7\3}+\chi^7\Ct^{\3}_{7\3}+\chi^8\Ct^{\3}_{8\3})\phi^3-
\Ct^{\3}_{\1\2}\phi^{1}\phi^{2}\ ,\\[4mm]
2\dot \chi^7=\Ct^7\chi^7 -\Ct^7_{1\1}|\phi^1|^2-\Ct^7_{2\2}|\phi^2|^2
+\Ct^7_{3\3}|\phi^3|^2\ ,\\[4mm]
2\dot \chi^8=\Ct^8+\Ct^7\chi^8-\Ct^8_{1\1}|\phi^1|^2-\Ct^8_{2\2}|\phi^2|^2
+\Ct^8_{3\3}|\phi^3|^2\ ,
\end{matrix}
\end{equation}
where $\Ct^7$ and $\Ct^8$ are given in (\ref{3.17a}). The superpotential $W$ becomes
\begin{multline}
	2 W = - \s_1^2 (\tilde{C}^{\bar{1}}_{2\bar{3}} + \tilde{C}^{\bar{1}}_{7\bar{1}}) |\phi^1|^2 - \s_2^2 (\tilde{C}^{\bar{2}}_{\bar{3} 1} + \tilde{C}^{\bar{2}}_{7\bar{2}}) |\phi^2|^2 - \s_3^2 (\tilde{C}^{\bar{3}}_{\bar{1}\bar{2}} - \tilde{C}^{\bar{3}}_{7\bar{3}}) |\phi^3|^2 \\ + \s_1 \s_2 \s_3 (\phi^1 \phi^2 \bar{\phi}^{\3} + \bar{\phi}^{\1} \bar{\phi}^{\2} \phi^3) + (\s_1^2 \Ct^{\1}_{7\1} |\phi^1|^2 + \s_2^2 \Ct^{\2}_{7\2} |\phi^2|^2 - \s_3^2 \Ct^{\3}_{7\3} |\phi^3|^2) \chi^7 \\ + (\s_1^2 \Ct^{\1}_{8\1} |\phi^1|^2 + \s_2^2 \Ct^{\2}_{8\2} |\phi^2|^2 - \s_3^2 \Ct^{\3}_{8\3} |\phi^3|^2) \chi^8 - \Ct^8 K_{8i} \chi^i - \frac{1}{2} \Ct^7 K_{ij} \chi^i \chi^j \; .
\end{multline}
where $K$ is the Killing metric $K(I,J) = - \tr(I,J)$ for the rescaled generators~(\ref{III}). The necessity to introduce $K$ is due to the fact that $\It_7$ and $\It_8$ are not mutually orthogonal for general values of $k$ and $l$. The explicit form of $K$ is given by
\begin{equation}
\begin{aligned}
	K_{\a\bb} &= -\tr(\It_\a \It_{\bb}) = \s^2_\a \delta_{\a\bb} \qquad \text{(no sum over $\a$)} \; , \\
	K_{77} &= -\tr(\It_7 \It_7) = \frac{8(k^2-kl+l^2)}{\mu^2 \Delta^2} \; , \\
	K_{88} &= -\tr(\It_8 \It_8) = \frac{8(k^2+kl+l^2)}{3\mu^2 \Delta^2} \; , \\
	K_{78} &= -\tr(\It_7 \It_8) = - \frac{4(k-l)(k+l)}{\sqrt{3}\mu^2 \Delta^2}
\end{aligned}
\end{equation}
with all other components vanishing. We are now in a position to express the first-order equations~(\ref{3.19}) in terms of the superpotential $W$,
\begin{equation}\label{insteqs_spot}
	\dot{\phi}^\a = - K^{\a\bb} \frac{\partial W}{\partial \bar{\phi}^{\bb}} \; , \qquad\qquad 
	\dot{\chi}^i = - K^{ij} \frac{\partial W}{\partial \chi^j} \; .
\end{equation}
The non-vanishing components of the inverse Killing metric are given by
\begin{equation}
	K^{\a\bb} = \s^{-2}_\a \delta^{\a\bb} \; , \quad
	K^{77} = - \frac{\mu^4}{4} K_{88} \; , \quad
	K^{88} = - \frac{\mu^4}{4} K_{77} \; , \quad
	K^{78} = \frac{\mu^4}{4} K_{78} \; ,
\end{equation}
such that $K^{\a\bb} K_{\g\bb} = \delta^\a_\g$, $K^{\a\bb} K_{\a\bar{\gamma}} = \delta^{\bb}_{\bar{\gamma}}$ and $K^{ij} K_{jk} = \delta^i_k$.

Eq.~(\ref{3.19}) is a complicated set of coupled, non-linear first-order ordinary differential equations and finding the general solution is a formidable task. Instead, one may consider simplifications of these equations by setting some of the fields to zero and hope to find explicit solutions for these special cases. Indeed, eq.~(\ref{3.19}) admits a particularly simple yet important special solution, namely
\begin{equation}
\begin{aligned}
 &\phi^1 = \phi^2 = \phi^3 = \chi^7 = 0 \; ,\\
 &\chi^8 (\tau) = \begin{cases}
                   -\frac{\tilde{C}^8}{\tilde{C}^7} + A\cdot\exp\left(\frac{\tilde{C}^7}{2}\,\tau \right) & \text{if $\tilde{C}^7 \neq 0$},\\
                   \frac{\tilde{C}^8}{2}\,\tau + B & \text{if $\tilde{C}^7 = 0$}.
                  \end{cases}
\end{aligned}
\end{equation}
where $A,B\in\mathbb{R}$ are constants of integration. For $\tilde{C}^7 \neq 0$, $A=0$, this solution is stationary and corresponds to the abelian (rescaled, if $\tilde{C}^8\neq 0$) canonical connection on a line bundle over $X_{k,l}$. This is arguably the simplest example for a $G_2$-instanton on Aloff-Wallach spaces. A similar conclusion also holds for $\tilde{C}^7 = \tilde{C}^8 = 0$ with the rescaled canonical connection corresponding to the case $B\neq 0$.

Before specialising to $k=l=1$, we briefly mention that the second-order equations of motion and the potential $V$ for the scalar fields can be obtained straightforwardly from the above first order equations by simply applying another time derivative to~(\ref{insteqs_spot}). The result can be written as
\begin{equation}\label{eom_pot}
	\ddot{\phi}^\a = K^{\a\bb} \frac{\partial V}{\partial \bar{\phi}^{\bb}} \; , \qquad\qquad 
	\ddot{\chi}^i = K^{ij} \frac{\partial V}{\partial \chi^j} \; .
\end{equation}
The potential $V$ is determined by the usual formula in terms of the superpotential
\begin{equation}\label{pot_spot}
	V = K^{\a\bb} W_\a W_{\bb} + \frac{1}{2} K^{ij} W_i W_j \; ,
\end{equation}
where we introduced the shorthand notation $W_\a = \partial W / \partial \phi^\a$, $W_{\bb} = \partial W / \partial \bar{\phi}^{\bb}$ and $W_i = \partial W / \partial \chi^i$. Computing the gradient of $V$ yields
\begin{equation}\label{pot_spot_grad}
	V_\a = K^{\b\bar\gamma} (W_{\a\b} W_{\bar\gamma} + W_{\a\bar\gamma} W_{\b}) + K^{ij} W_{\a i} W_j \; , \quad
	V_i = K^{\a\bb} (W_{i\a} W_{\bb} + W_{i\bb} W_{\a}) + K^{jk} W_{ij} W_k \; .
\end{equation}
From this and~(\ref{pot_spot}) we can read off that critical points of the superpotential are both zeros and critical points of the potential. On the other hand, the critical points of $V$ fall into two categories: zero-energy ones ($V=0$) and positive-energy ones ($V>0$). The former are precisely the critical points of $W$, which will be studied further for the special case $k=l=1$ in the remainder of this section. However, the positive-energy critical points of $V$ do not correspond to critical points of $W$. Instead, for them the gradient of $W$ is a ``zero eigenvector'' of the Hessian of $W$. They will not play a role in our analysis.

\bigskip

\noindent
{\bf Specialization to $k=l=1$.}
For the special case of $X_{1,1}$, with $\s_\a$ and $\mu$ given in (\ref{sss}), from (\ref{3.19}) we obtain
\begin{equation}\label{3.20}
\begin{aligned}
2\dot\phi^1 &= (\a -1 + \chi^7 - \sqrt{3}\,\chi^8)\phi^1-\a\,\bar\phi^{\2}\phi^3\ ,\\[1.5mm]
2\dot\phi^2 &= (\a -1 + \chi^7 + \sqrt{3}\,\chi^8)\phi^2-\a\,\bar\phi^{\1}\phi^3\ ,\\[1.5mm]
 \dot\phi^3 &= (\a +1 - \chi^7)\phi^3-\a\,\phi^1\phi^2\ ,\\[1.5mm]
2\dot \chi^7 &= -\chi^7+|\phi^1|^2+|\phi^2|^2-|\phi^3|^2\ ,\\[1.5mm]
2\dot \chi^8 &= -\chi^8-\sqrt{3}\,|\phi^1|^2+\sqrt{3}\,|\phi^2|^2\ ,
\end{aligned}
\end{equation}
with $\a=\pm 1$ for the 3-Sasakian structure on $X_{1,1}$. The Killing metric in this case becomes diagonal with non-zero components
\begin{equation}
	K_{1\1} = K_{2\2} = 2 K_{3\3} = K_{77} = K_{88} = 2 \; .
\end{equation}
The superpotential simplifies to
\begin{multline}\label{spot11}
	W = (1 - \a) \left( |\phi^1|^2 + |\phi^2|^2 \right) - (1 + \a) |\phi^3|^2 + \frac{1}{2} \left((\chi^7)^2 + (\chi^8)^2 \right) \\ + \a \left( \phi^1 \phi^2 \bar{\phi}^{\3} + \bar{\phi}^{\1} \bar{\phi}^{\2} \phi^3 \right) - \left( |\phi^1|^2 + |\phi^2|^2 - |\phi^3|^2 \right)\chi^7 + \sqrt{3}\left( |\phi^1|^2 - |\phi^2|^2 \right)\chi^8
\end{multline}
and (\ref{3.20}) may be written as
\begin{equation}
	2\dot\phi^1 = - W_\1 \; , \quad
	2\dot\phi^2 = - W_\2 \; , \quad
	2\dot\phi^3 = - 2W_\3 \; , \quad
	2\dot\chi^7 = - W_7 \; , \quad
	2\dot\chi^8 = - W_8 \; .
\end{equation}
The superpotential~(\ref{spot11}) is invariant under global U(1)$\times$U(1) transformations of the form
\begin{equation}\label{phasesU1xU1}
	(\phi^1, \phi^2, \phi^3) \;\;\mapsto\;\; (e^{i\delta_1}\phi^1, e^{i\delta_2}\phi^2, e^{i\delta_3}\phi^3) \qquad
	\text{with}\; \delta_1 + \delta_2 - \delta_3 = 0 \mod 2\pi \; .
\end{equation}
Note that the phases of the $\phi^\a$ only enter in the cubic terms $\left( \phi^1 \phi^2 \bar{\phi}^{\3} + \bar{\phi}^{\1} \bar{\phi}^{\2} \phi^3 \right)$ in the superpotential, which are thus proportional to $\cos(\arg\phi^1 + \arg\phi^2 - \arg\phi^3)$. The superpotential is extremised when $\arg\phi^1 + \arg\phi^2 - \arg\phi^3 = 0$ or $\pi$ and, together with~(\ref{phasesU1xU1}), this allows us to consider purely real fields when searching for extrema of $W$. After fixing $\phi^\a \in \mathbb{R}$, there is a residual symmetry which acts by flipping the sign of any two of the three complex functions $\phi^\a$. Therefore, we can restrict ourselves not only to real fields but also take, for example, $\phi^1$ and $\phi^2$ non-negative when searching for extrema of $W$.

In addition, there is a $Z_2$-symmetry which acts by interchanging $\phi^1$ and $\phi^2$ accompanied by a sign flip of $\chi^8$
\begin{equation}\label{Z2_symm}
	(\phi^1, \phi^2, \chi^8) \;\;\mapsto\;\; (\phi^2, \phi^1, -\chi^8) \; .
\end{equation}

\bigskip

\noindent
{\bf Explicit solutions for $k=l=1$.}
We will begin by finding the extrema of the superpotential~(\ref{spot11}). Making use of the argument given at the end of the previous section, we take all fields to be real and $\phi^1$, $\phi^2$ non-negative. We then need to solve the following equations in five real variables
\begin{equation}\label{11_crit_pt_alg_eqs}
\begin{aligned}
(\a -1 + \chi^7 - \sqrt{3}\,\chi^8)\phi^1-(\pm)\a\,\phi^2\phi^3 &= 0\ ,\\[1.5mm]
(\a -1 + \chi^7 + \sqrt{3}\,\chi^8)\phi^2-(\pm)\a\,\phi^1\phi^3 &= 0\ ,\\[1.5mm]
(\a +1 - \chi^7)\phi^3-(\pm)\a\,\phi^1\phi^2 &= 0\ ,\\[1.5mm]
-\chi^7+(\phi^1)^2+(\phi^2)^2-(\phi^3)^2 &= 0\ ,\\[1.5mm]
-\chi^8-\sqrt{3}\,(\phi^1)^2+\sqrt{3}\,(\phi^2)^2 &= 0\ .
\end{aligned}
\end{equation}
The sign ambiguity in the first three equations is a consequence of $\cos(\arg\phi^1 + \arg\phi^2 - \arg\phi^3) = \pm 1$ at the extrema. The last two equations may be used to immediately eliminate $\chi^7$ and $\chi^8$ and one is then left with three cubic equations for three unknowns,
\begin{equation}\label{11_crit_pt_alg_eqs_chis_elim}
\begin{aligned}
(\a -1 + 4(\phi^1)^2 - 2(\phi^2)^2 - (\phi^3)^2)\phi^1-(\pm)\a\,\phi^2\phi^3 &= 0\ ,\\[1.5mm]
(\a -1 + 4(\phi^2)^2 - 2(\phi^1)^2 - (\phi^3)^2)\phi^2-(\pm)\a\,\phi^1\phi^3 &= 0\ ,\\[1.5mm]
(\a +1 - (\phi^1)^2 - (\phi^2)^2 + (\phi^3)^2)\phi^3-(\pm)\a\,\phi^1\phi^2 &= 0\ .
\end{aligned}
\end{equation}

One obvious solution is $\phi^\a = \chi^i = 0$. However, the full analysis depends on the choice of $\a = +1$ or $-1$. The results for $\a = +1$ are summarised in the following table:
\begin{center}
\begin{tabular}[h]{|c|c|c|c|c|c|c|}\hline
$\phi^1$	&	$\phi^2$	&	$\phi^3$	&	$\chi^7$	&	$\chi^8$ &	Eigenvalues of Hessian & $W\vphantom{\frac{\frac{1}{2}}{2}}$ \\\hline
$0$	& $0$	&	$0$	&	$0$	&	$0$	&	$(-,+,+,0,0)$ & $0$ \\
$1$	& $1$	&	$\pm 1$	&	$1$	&	$0$	&	$(-,-,+,+,+)$ & $-1/2$ \\
\hline
\end{tabular}
\end{center}
where the sign ambiguity in the $\phi^3$ column stems from the fact that $\cos(\arg\phi^1 + \arg\phi^2 - \arg\phi^3) = \pm 1$ at the extrema (cf. eq.~(\ref{11_crit_pt_alg_eqs})). There is one saddle point and a degenerate critical point at the origin. The appearance of the degenerate critical point can be understood from a physics perspective by noticing that $\phi^1$ and $\phi^2$ are massless and therefore correspond to flat directions in the space of solutions.

The results for $\a = -1$ are summarised in the table below:
\begin{center}
\begin{tabular}[h]{|c|c|c|c|c|c|c|c|}\hline
$\phi^1$	&	$\phi^2$	&	$\phi^3$	&	$\chi^7$	&	$\chi^8$ &	Eigenvalues of Hessian & $W\vphantom{\frac{\frac{1}{2}}{2}}$ \\\hline
$0$	& $0$	&	$0$	&	$0$	&	$0$	&	$(+,+,+,+,0)$ & $0$ \\
$1/\sqrt{2}$ & $0$ &	$0$	&	$1/2$	&	$-\sqrt{3}/2$ &	$(-,+,+,+,+)$ & $1/2$ \\
$c_+$	& $c_-$ & 	$\pm 2/3$&	$1/6$	&	$-\sqrt{35}/6$ & $(-,-,+,+,+)$ & $31/54$ \\
$2\sqrt{2}/3$ & $2\sqrt{2}/3$ & $\pm 2/3$ & $4/3$ &	$0$	&	$(-,-,+,+,+)$ & $40/27$ \\
$1$	& $1$	&	$\pm 1$	&	$1$	&	$0$	&	$(-,-,-,+,+)$ & $3/2$ \\
\hline
\end{tabular}
\end{center}
where $c_\pm = \frac{1}{6}\sqrt{11 \pm\sqrt{105}}$, and again the sign ambiguity in the $\phi^3$ column is due to the fact that $\cos(\arg\phi^1 + \arg\phi^2 - \arg\phi^3) = \pm 1$ at the extrema. The origin is a degenerate critical point. The field $\phi^3$ is massless and hence corresponds to a flat direction in the space of solutions, which explains why the critical point at the origin is degenerate. In addition, there are four isolated saddle points.

A few remarks are in order concerning the critical points found above. First of all, we note that the critical point at the origin (i.e. where all scalar fields vanish) corresponds to the abelian canonical connection on a rank-3 complex vector bundle over $X_{1,1}$ and is thus arguably the simplest explicit example of a $G_2$-instanton on an Aloff-Wallach space. Also, the point where $\chi^8=0$ and all other scalar fields are equal to unity corresponds to a flat connection $\Fcal = 0$. These observations are valid for both choices of $\a = +1$ or $-1$.

Now that we have found the critical points of the superpotential, we consider the gradient flow connecting suitable pairs of them. In other words, we look for solutions of~(\ref{3.20}), which start at $\tau=-\infty$ from a critical point with a larger value of $W$ and flow as $\tau\rightarrow\infty$ towards a critical point with a smaller value of $W$. These kink configurations are finite-action solutions of~(\ref{3.5}) and thus allow a physical interpretation as Spin(7)-instantons on $\R\times X_{1,1}$.

In the search for instanton solutions, one is immediately faced with two technical difficulties. First, the structure of the equations which need to be solved is such that conventional analytic methods (and known exact solution ans\"atze) are not applicable. For example, the well-known hyperbolic tangent type kink solutions, which \emph{inter alia} work in one dimension lower~\cite{13}, do not respect the structure of~(\ref{3.20}). This means we need to resort to numerical methods.

Second, with the exception of the degenerate critical point at the origin, all other critical points are isolated saddle points. Solutions flowing towards these points are unstable. For a given starting point, there is exactly one trajectory whose end point is an isolated saddle point and it is crucial to pick the initial direction to be exactly along this unique trajectory. Combined with the first point, this presents us with a numerical ``fine-tuning problem'' when it comes to choosing the correct initial conditions for the desired flow. One would somehow need to know the trajectory's direction at the starting point before even attempting to (numerically) solve the equations, leaving oneself with a ``fishing in the dark'' situation. Moreover, even the smallest deviation from the correct direction will lead to solutions which, instead of approaching the saddle point, will roll off to $\pm\infty$ swiftly.

\begin{figure}[t]
	\centering
	\includegraphics[width=0.75\textwidth]{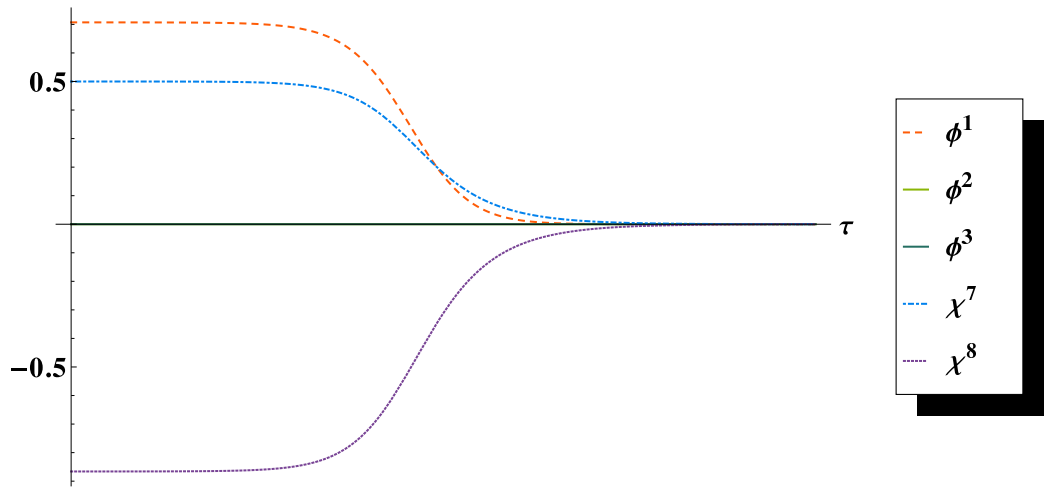}
	\caption{Kink solution for $\a = -1$ flowing from $W = 1/2$ at $\tau=-\infty$ to $W = 0$ as $\tau\rightarrow\infty$. $\phi^2$ and $\phi^3$ are zero everywhere and thus their plot coincides with the $\tau$-axis.}
	\label{flow_eqs_minussqrt2_ex1}
\end{figure}

There is only one case where this ``fine-tuning problem'' does not occur and where we have been able to find an explicit (numerical) solution. It is the kink solution for $\a = -1$ flowing from $W = 1/2$ at $\tau=-\infty$ to $W = 0$ as $\tau\rightarrow\infty$. The numerical solution for this case is shown in figure~\ref{flow_eqs_minussqrt2_ex1}.
It should be noted that the shape of these curves resembles that of a hyperbolic tangent type kink. Indeed, although a hyperbolic tangent ansatz does not solve equations~(\ref{3.20}), it does provide a good approximation. The maximal deviation from the actual numerical solution is of the order of $1 \%$.

\bigskip

\noindent
{\bf Acknowledgements}

\noindent 
We thank Derek Harland for helpful comments.
This work was supported in part by the cluster of excellence EXC 201 ``Quantum Engineering and Space-Time Research'', by the Deutsche Forschungsgemeinschaft (DFG), by the Russian Foundation for Basic Research (grant RFBR 09-02-91347) and by the Heisenberg-Landau program.



\bigskip


\begin{thebibliography}{99}

\bibitem{1}
  M.B.~Green, J.H.~Schwarz and E.~Witten,
  {\it Superstring theory},\\
  Cambridge University Press, Cambridge, 1987.

\bibitem{2}
  E.~Corrigan, C.~Devchand, D.B.~Fairlie and J.~Nuyts,
  ``First order equations for gauge fields in spaces of dimension
    greater than four,''
  Nucl.\ Phys.\ B {\bf 214} (1983) 452.

\bibitem{3}
  R.S.~Ward,
  ``Completely solvable gauge field equations in dimension
    greater than four,''\\
  Nucl.\ Phys.\ B {\bf 236} (1984) 381.

\bibitem{4}
  S.K.~Donaldson,
  ``Anti-self-dual Yang-Mills connections on a complex algebraic surface
    and stable vector bundles,''
  Proc.\ Lond.\ Math.\ Soc.\ {\bf 50} (1985) 1;

  S.K.~Donaldson,
  ``Infinite determinants, stable bundles and curvature,''\\
  Duke Math.\ J.\ {\bf 54} (1987) 231;

  K.K.~Uhlenbeck and S.-T.~Yau,
  ``On the existence of Hermitian-Yang-Mills connections on stable bundles
    over compact K\"ahler manifolds,''
  Commun.\ Pure Appl.\ Math.\ {\bf 39} (1986) 257.
  
\bibitem{5}
  M.~Mamone~Capria and S.M.~Salamon,
  ``Yang-Mills fields on quaternionic spaces,''\\
  Nonlinearity {\bf 1} (1988) 517;

  R.~Reyes~Carri\'on,
  ``A generalization of the notion of instanton,''
  Diff.\ Geom.\ Appl.\ {\bf 8} (1998) 1.

\bibitem{6}
  L.~Baulieu, H.~Kanno and I.M.~Singer,
  ``Special quantum field theories in eight and other dimensions,''
  Commun.\ Math.\ Phys.\ {\bf 194} (1998) 149
  [arXiv:hep-th/9704167].

\bibitem{7}
  G.~Tian,
  ``Gauge theory and calibrated geometry,''\\
  Ann.\ Math.\ {\bf 151} (2000) 193
  [arXiv:math/0010015 [math.DG]];

  T.~Tao and G.~Tian,
  ``A singularity removal theorem for Yang-Mills fields in higher dimensions,''
  J.\ Amer.\ Math.\ Soc.\ {\bf 17} (2004) 557.

\bibitem{8}
  S.K.~Donaldson and R.P.~Thomas,
  ``Gauge theory in higher dimensions,''\\
  in: {\it The Geometric Universe},
  Oxford University Press, Oxford, 1998;

  S.K.~Donaldson and E.~Segal,
  ``Gauge theory in higher dimensions II'',\\
  arXiv:0902.3239 [math.DG].

\bibitem{9}
  A.D.~Popov,
  ``Non-Abelian vortices, super-Yang-Mills theory and Spin(7)-instantons,''\\
  Lett.\ Math.\ Phys.\ {\bf 92} (2010) 253
  [arXiv:0908.3055 [hep-th]];

  D.~Harland and A.D.~Popov,
  ``Yang-Mills fields in flux compactifications on homogeneous manifolds
  with SU(4)-structure,''
  arXiv:1005.2837 [hep-th];

  A.D.~Popov and R.J.~Szabo,
  ``Double quiver gauge theory and nearly K\"ahler flux compactifications,''
  arXiv:1009.3208 [hep-th].

\bibitem{10}
  D.B.~Fairlie and J.~Nuyts,
  ``Spherically symmetric solutions of gauge theories in eight dimensions,''
  J.\ Phys.\ A {\bf 17} (1984) 2867;

  S.~Fubini and H.~Nicolai,
  ``The octonionic instanton,''
  Phys.\ Lett.\ B {\bf 155} (1985) 369;

  T.A.~Ivanova and A.D.~Popov,
  ``Self-dual Yang-Mills fields in $d{=}7, 8$, octonions and Ward equations,''
  Lett.\ Math.\ Phys.\  {\bf 24} (1992) 85;

  T.A.~Ivanova and A.D.~Popov,
  ``(Anti)self-dual gauge fields in dimension $d{\ge}4$,''\\
  Theor.\ Math.\ Phys.\ {\bf 94} (1993) 225.

  M.~G\"unaydin and H.~Nicolai,
  ``Seven-dimensional octonionic Yang-Mills instanton and its extension to an
  heterotic string soliton,''
  Phys.\ Lett.\  B {\bf 351} (1995) 169
  [arXiv:hep-th/9502009].

\bibitem{11}
  O.~Lechtenfeld, A.D.~Popov and R.J.~Szabo,
  ``Noncommutative instantons in higher dimensions, vortices and  topological
  K-cycles,''
  JHEP {\bf 12} (2003) 022
  [arXiv:hep-th/0310267];

  A.D.~Popov, A.G.~Sergeev and M.~Wolf,
  ``Seiberg-Witten monopole equations on noncommutative $\R^4$,''
  J.\ Math.\ Phys.\  {\bf 44} (2003) 4527
  [arXiv:hep-th/0304263];

  O.~Lechtenfeld, A.D.~Popov and R.J.~Szabo,
  ``SU(3)-equivariant quiver gauge theories and nonabelian vortices,''
  JHEP {\bf 08} (2008) 093
  [arXiv:0806.2791 [hep-th]].

\bibitem{12}
  T.A.~Ivanova, O.~Lechtenfeld, A.D.~Popov and T.~Rahn,
  ``Instantons and Yang-Mills flows on coset spaces,''
  Lett.\ Math.\ Phys.\  {\bf 89} (2009) 231
  [arXiv:0904.0654 [hep-th]];

  T.~Rahn,
  ``Yang-Mills equations of motion for the Higgs sector of SU(3)-equivariant
    quiver gauge theories,''
  J.\ Math.\ Phys.\  {\bf 51} (2010) 072302
  [arXiv:0908.4275 [hep-th]].

\bibitem{13}
  D.~Harland, T.A.~Ivanova, O.~Lechtenfeld and A.D.~Popov,\\
  ``Yang-Mills flows on nearly K\"ahler manifolds and $G_2$-instantons,''\\
  Commun.\ Math.\ Phys.\  {\bf 300} (2010) 185
  [arXiv:0909.2730 [hep-th]];

  I.~Bauer, T.A.~Ivanova, O.~Lechtenfeld and F.~Lubbe,
  ``Yang-Mills instantons and dyons on homogeneous $G_2$-manifolds,''
  JHEP {\bf 10} (2010) 044
  [arXiv:1006.2388 [hep-th]].

\bibitem{14}
  H.N.~S\`a Earp,
  ``Instantons on $G_2$-manifolds'',
  PhD thesis, Imperial College London, 2009.

\bibitem{15}
  C.~Lewis,  ``Spin(7) instantons'',
  PhD thesis, Oxford University, 1998. 

\bibitem{16}
  M.~Grana,
  ``Flux compactifications in string theory: A comprehensive review,''\\
  Phys.\ Rept.\ {\bf 423} (2006) 91
  [arXiv:hep-th/0509003];

  M.R.~Douglas and S.~Kachru,
  ``Flux compactification,''\\
  Rev.\ Mod.\ Phys.\ {\bf 79} (2007) 733
  [arXiv:hep-th/0610102];

  R.~Blumenhagen, B.~Kors, D.~L\"ust and S.~Stieberger,
  ``Four-dimensional string compactifications with D-branes, orientifolds
    and fluxes,''
  Phys.\ Rept.\ {\bf 445} (2007) 1
  [arXiv:hep-th/0610327].

\bibitem{17}
  A.~Strominger,
  ``Superstrings with torsion,''
  Nucl.\ Phys.\ B {\bf 274} (1986) 253;

  C.M.~Hull,
  ``Anomalies, ambiguities and superstrings,''
  Phys.\ Lett.\ B {\bf 167} (1986) 51 (1986);

  C.M.~Hull,
  ``Compactifications of the heterotic superstring,''\\
  Phys.\ Lett.\ B {\bf 178} (1986) 357 (1986);

  D.~L\"ust,
  ``Compactification of ten-dimensional superstring theories over Ricci flat
    coset spaces,''\\
  Nucl.\ Phys.\ B {\bf 276} (1986) 220;

  B.~de Wit, D.J.~Smit and N.D.~Hari Dass,
  ``Residual supersymmetry of compactified D=10 supergravity,''
  Nucl.\ Phys.\ B {\bf 283} (1987) 165.

\bibitem{18}
  S.~Aloff and N.~Wallach, 
  ``An infinite family of distinct 7-manifolds admitting positively curved 
  Riemannian structures,'' 
  Bull.\ Amer.\ Math.\ Soc.\ {\bf 81} (1975) 93.

\bibitem{19} 
  F.M.~Cabrera, M.D.~Monar and A.F.~Swann, 
  ``Classification of $G_2$-structures,'' \\
  J.\ London Math.\ Soc.\ {\bf 53} (1996) 407;

  Th.~Friedrich, I.~Kath, A.~Moroianu and U.~Semmelmann, 
  ``On nearly parallel $G_2$-structures,'' 
  J.\ Geom.\ Phys.\ {\bf 23} (1997) 259;

  Hong Van Le and M.~Munir, 
  ``Classification of compact homogeneous spaces with invariant $G_2$-structures,''
  arXiv:0912.0169 [math.DG].

\bibitem{20}
  M.~Cvetic, G.W.~Gibbons, H.~Lu and C.N.~Pope,
  ``Hyper-K\"ahler Calabi metrics, L$^2$ harmonic forms, resolved M2-branes, and
  AdS$_4$/CFT$_3$ correspondence,''\\
  Nucl.\ Phys.\ B {\bf 617} (2001) 151
  [arXiv:hep-th/0102185];

  M.~Cvetic, G.W.~Gibbons, H.~Lu and C.N.~Pope,
  ``Cohomogeneity one manifolds of Spin(7) and $G_2$ holonomy,''
  Phys.\ Rev.\ D {\bf 65} (2002) 106004
  [arXiv:hep-th/0108245];

  Y.~Konishi and M.~Naka,
  ``Coset construction of Spin(7), $G_2$ gravitational instantons,''\\
  Class.\ Quant.\ Grav.\ {\bf 18} (2001) 5521
  [arXiv:hep-th/0104208].

\bibitem{21}
  H.~Kanno and Y.~Yasui,
  ``On Spin(7) holonomy metric based on SU(3)/U(1),''\\
  J.\ Geom.\ Phys.\ {\bf 43} (2002) 293
  [arXiv:hep-th/0108226];

  H.~Kanno and Y.~Yasui,
  ``On Spin(7) holonomy metric based on SU(3)/U(1). II,''\\
  J.\ Geom.\ Phys.\ {\bf 43} (2002) 310
  [arXiv:hep-th/0111198];

  A.~Bilal, J.P.~Derendinger and K.~Sfetsos,
  ``(Weak) $G_2$ holonomy from self-duality, flux and supersymmetry,''
  Nucl.\ Phys.\ B {\bf 628} (2002) 112
  [arXiv:hep-th/0111274].

\bibitem{22}
  S.~Gukov and J.~Sparks,
  ``M-theory on Spin(7) manifolds. I,''\\
  Nucl.\ Phys.\ B {\bf 625} (2002) 3
  [arXiv:hep-th/0109025];

  G.~Curio, B.~Kors and D.~L\"ust,
  ``Fluxes and branes in type II vacua and M-theory geometry with $G_2$ and
  Spin(7) holonomy,''
  Nucl.\ Phys.\ B {\bf 636} (2002) 197
  [arXiv:hep-th/0111165].

\bibitem{23}
  A.D.~Popov,
  ``Hermitian-Yang-Mills equations and pseudo-holomorphic bundles on nearly
  K\"ahler and nearly Calabi-Yau twistor 6-manifolds,''\\
  Nucl.\ Phys.\ B {\bf 828} (2010) 594
  [arXiv:0907.0106 [hep-th]].

\end{thebibliography}
\end{document}